\documentclass[5p,times]{elsarticle}
\usepackage[utf8]{inputenc}

\newtheorem{problem}{Problem}
\newcommand{\eps}{\varepsilon}
\newcommand\tab[1][1cm]{\hspace*{#1}}
\usepackage{caption}
\usepackage{color}
\usepackage{natbib}
\usepackage{subcaption}
\usepackage[]{graphicx} 

\usepackage{amsmath}
\usepackage{algorithm}
\usepackage[noend]{algpseudocode}

\definecolor{ForestGreen}{rgb}{0.3,0.7,0.3}

\newcommand{\mpch}{Mpc/$h$}

\begin{document}
\title{Scalable Streaming Tools for Analyzing $N$-body Simulations:\\ Finding Halos and Investigating Excursion Sets in One Pass}
\author[1]{Nikita~Ivkin \fnref{ANL} }
                                            \ead{nivkin1@jhu.edu}
\author[1]{Zaoxing~Liu \fnref{ANL} }
\author[2]{Lin~F.~Yang \fnref{ANL} }
\author[1]{Srinivas~Suresh~Kumar \fnref{GTS} \corref{}}
\author[1]{Gerard~Lemson \fnref{GTS} \corref{}}
\author[3]{Mark~Neyrinck \fnref{M}} 
\author[1]{Alexander~S.~Szalay}
\author[1]{Vladimir~Braverman \fnref{V}}
\author[1]{Tamas~Budavari \fnref{GTS} \corref{}}
\address[1]{Johns Hopkins Universitye}
\address[2]{Princeton University}
\address[3]{Institute for Computational Cosmology, Durham University}

\fntext[ANL]{This material is based upon work supported in part by the NSF Grants IIS- 1447639, EAGER CCF- 1650041}
\fntext[GTS]{This material is based upon work supported in part by the NSF Grants IIS- 1447639}
\fntext[M]{Thanks the UK Science and Technology Facilities Council (ST/L00075X/1) for financial support}
\fntext[V]{This material is based upon work supported in part by the National Science Foundation under Grants No. 1447639, 1650041 and 1652257, Cisco faculty award, and by the ONR Award N00014-18-1-2364}

\date{}
\begin{abstract}
Cosmological $N$-body simulations play a vital role in studying models for the evolution of the Universe. To compare to observations and make a scientific inference, statistic analysis on large simulation datasets, e.g., finding halos, obtaining multi-point correlation functions, is crucial. However, traditional in-memory methods for these tasks do not scale to the datasets that are forbiddingly large in modern simulations.
Our prior paper~\cite{liu2015streaming} proposes memory-efficient streaming algorithms that can find the largest halos in a simulation with up to $10^9$ particles on a small server or desktop. However, this approach fails when directly scaling to larger datasets. This paper presents a robust streaming tool that leverages state-of-the-art techniques on GPU boosting, sampling, and parallel I/O, to significantly improve performance and scalability. 
Our rigorous analysis of the sketch parameters improves the previous results from finding the centers of the $10^3$ largest  halos~\cite{liu2015streaming} to $\sim 10^4-10^5$, and reveals the trade-offs between memory, running time and number of halos. Our experiments show that our tool can scale to datasets with up to $\sim 10^{12}$ particles while using less than an hour of running time on a single GPU Nvidia GTX 1080. 
\end{abstract}
\maketitle


\section{Introduction}
Cosmology is a field in physics and astrophysics that focuses on the study of the large-scale distribution of matter in the universe. Advanced computer simulations have become essential tools for understanding how matter organizes itself in galaxies, clusters of galaxies and large-scale structures (e.g., \cite{springel2005cosmological}). Many such simulations operate with a set of particles in a fixed cubic volume: at each step of the simulation the gravitational force field is computed and the velocities and positions of particles are recomputed according to that force field. Running large-scale simulations of this type is very expensive in terms of computational resources, both in running time and memory. Additionally, even if the simulation results are computed, the analysis of the dataset requires resources that are usually beyond the capabilities of many researchers. For example, to host a single snapshot of a simulation with roughly a trillion particles (e.g., \cite{angulo2012scaling, potter2017pkdgrav3}) requires tens of Terabytes of memory. Storing such a large number of particles is not only expensive but also challenging. 

One of the essential steps in the analysis of these simulations is the identification of ``halos'' \cite{knebe2013structure}, which are concentrations of mass.
Galaxies are expected to form inside these halos. 
Finding halos in the output of the simulation allows astronomers to compute important statistics, e.g., mass functions \cite{karttunen2016fundamental}. 
These statistics are crucial for comparison between theories and observations. 
Although from an astronomical perspective the concept of a ``halo'' is fairly well understood, the mathematical definition of halos in a simulation varies among simulation and analysis methods.  
For instance, \cite{planelles2010asohf} defines it as mass blobs around the density peaks above some thresholds;
\cite{davis1985evolution} defines it as the connected components of the distances graph on the particles. A definition that does not use the density, instead uses particle crossings \cite{falck2012origami}.
The lack of agreement upon a single definition of a halo makes it difficult to uniquely compare the results of different halo-finding algorithms. 
Nevertheless, \cite{knebe2011haloes} evaluated 17 different algorithms and compared them using various criteria, and found broad agreement between them, but with many differences in detail for ambiguous cases.

Although there are a large number of algorithms and implementations \cite{knollmann2009ahf,gill2004evolution,planelles2010asohf,klypin1997particle,sutter2010examining,white2001halo,neyrinck2005voboz,davis1985evolution}, these approaches generally require the dataset to be stored entirely in memory.
Thus for state-of-the-art simulations \cite{angulo2012scaling, potter2017pkdgrav3}, which reach hundreds of billions and even over a trillion of particles, post-processing analysis becomes unfeasible unless using supercomputers of the same size that created the simulations in the first place. 
Recently in \cite{liu2015streaming} an approach was proposed that attacks the problem using solutions developed in the field of streaming algorithms.
Typical applications of streaming algorithms are very large datasets, where access to the data is restricted to be sequential and the working memory is much smaller than the dataset size. 
In their most restricted versions, streaming algorithms are supposed to make a single pass over the data using a polylogarithmic amount of memory in terms of the length of the sequence. 
Such restrictions force streaming algorithms to be randomized, providing approximate answers for problems of interest. 
Streaming algorithms have numerous applications in networking \cite{Zhang:2004:OIH,Lall:2006:DSA,Zhao:2007:DSA,univmon2016}, machine learning\cite{Beringer:2007:EIL,liberty2013simple}, and databases \cite{Rusu:2007:SAS,Spiegel:2006:GSR}. 
For a detailed review please refer to \cite{muthukrishnan2005data}. 

In \cite{liu2015streaming}, we developed a solution using the {\it Count Sketch} streaming algorithm \cite{charikar2002finding} to the halo finding problem, and presented the first results on how to find the top-k ($k\approx 1000$) largest halos in a dark matter simulation. 
In that paper, all experiments were running on relatively small data streams with at most $10^9$ items. 
One of the reasons for that was the rather poor {\it time} performance of the underlying algorithms, which would cause every experiment to take more than week to run. In this paper, we improve the implementation and push the number of halos' centers to be found to $\sim 10^4 - 10^5$. 

Our tool needs less than $5$ minutes to find the top $3\cdot 10^5$ heavy cells on a dataset with $10^{10}$ particles. Compared to previous results \cite{liu2015streaming}, which required more than 8 hours, it's more than a  $100\times$ improvement. This dataset consists of a snapshot of the Millennium dataset~\cite{springel2005cosmological} and we use a grid of $10^{11}$ cells in our algorithm for approximation of the density field, which can be used further for astrophysical analysis. We port the entire Count-Sketch infrastructure into the GPU and thus make the tool significantly outperform the previous approach. 
In our analysis, we carefully investigate the trade-off between memory and the quality of the result.

In \cite{liu2015streaming} authors reduced the halo-finding problem to the problem of finding the top-$k$ densest cells in a regular mesh. This reduction shows that these densest cells are closely related to the space with the heaviest halos. 
In this paper, we consider another possible application, that of determining statistics on ``excursion sets''.
Kaiser \cite{kaiser1984spatial} investigated the clustering properties of the regions with a density higher than the average in Gaussian random fields. He showed that such regions cluster more strongly than those with lower over-densities and the strength of this effect increases with the density threshold. He used this as an explanation of the observed stronger clustering of galaxy clusters compared to the clustering of the galaxy distribution itself. Bardeen et al.~\cite{bardeen1986statistics}, refined this argument, focusing on the peaks of the density fields---the locations where galaxies and clusters are expected to form. 

This biased clustering phenomenon can be examined in an evolved density field by filtering regions in the dark matter distribution field, based on their density. This is equivalent to examining the ``heavy hitters'' in the counts-in-cells. We expect the randomized algorithms to not be exact, and it is interesting to investigate how this affects the clustering measure.

The outline of this paper is as the following. 
\begin{itemize}
\item In Section \ref{sec:streaming}, we formally define the structure of the streaming model in different settings, investigate the heavy hitter problem and its connection to the spatial statistics in $N$-body simulations, and describe the algorithms that are capable of solving the problem. 

\item In Section~\ref{sec:implementation}, we describe our implementation.We outline how successive improvements, in particular, the extensive usage of GPUs, make it possible to run our experiments in about 5 minutes, whereas in the previous paper it took 8 hours.

\item In Section~\ref{sec:evaluation}, we evaluate the accuracy of the results by comparing the streaming algorithm results with the exact results where possible. 
We do the comparison not only in the information-theoretical setting but also in the statistical setting, which is of astrophysical interest. 
Evaluations show that the approximate results accurately reproduce exact statistics. 
\item In Section~\ref{sec:conclusion}, we conclude the paper and discuss future work. 
\end{itemize}

\section{Methodology}
\label{sec:streaming}
In this section, we introduce our methods for efficiently analyzing cosmological datasets. First, we introduce the concept of streaming and explain how the problem of estimating density statistics can be approached from the perspective of finding frequent items in the stream. Then we recap the general idea and several crucial details of the heavy hitter algorithm named Count Sketch\cite{charikar2002finding}. 

\subsection{Streaming Model}
The streaming model was first formally introduced in the seminal paper \cite{alon1996space}. In this model, an algorithm is required to compute a certain function $f$ by making a single (or a few) pass(es) on a long stream of data $S = \{s_1, \ldots, s_m \}$ with a limited amount of memory. Elements of the stream $s_i$ are in an arbitrary order and belong to some given dictionary $D = \{d_1, \ldots, d_n\}$, where $d_j$ can represent different entities, such as integers, edges in a graph, sets, or rows of a matrix. For simplicity we will consider a dictionary of integers $D = [n] = \{1,\ldots, n\}$. Typically, both $m$ and $n$ are very large numbers such that it is usually infeasible to store the entire stream or even the frequency of each element in the dictionary. Thus in the streaming model aims for algorithms with very low memory usage, e.g. $o(n + m)$ bits. 
Due to such strong limitations, most streaming algorithms are randomized and have approximation error.


In this paper, we work on two cosmological $N$-body simulations with $10^{10}$ and $3\times 10^{11}$ particles,  respectively, resulting in several terabytes of data. 
In this setting, typical approaches for finding halos that require loading data into memory become inapplicable on common computing devices (e.g. laptop, desktop, or small workstations) for post-processing and analysis. 
In contrast, a streaming approach makes the analysis of such datasets feasible even on a desktop by lowering the memory footprint from several terabytes to less than a gigabyte. 

\begin{figure}[h]
            \includegraphics[width=8cm]{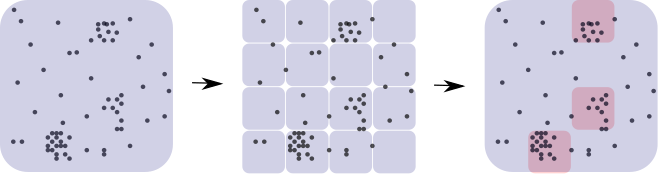}
            \centering
            \caption{Finding approximate dense areas with the help of a regular mesh and a streaming solution for finding the top $k$ most frequent items in the stream.}
            \label{fig:mesh}
\end{figure}

Much of the analysis of cosmological $N$-body simulations focuses on regions with a high concentration of particles. 
By putting a regular mesh on the simulation box, we can replace each particle with the ID of the cell it belongs to. 
Then using streaming algorithms we can find the $k$ most frequent cells, i.e. cells with the largest number of particles (see Figure \ref{fig:mesh}). 
Such statistics are very useful for analyzing a spatial distribution of particles on each iteration of the simulation, as shown in \cite{liu2015streaming} and as we will show in the current paper. 
One might think that this approach is too naive and just keeping a counter for each cell would provide the exact solution with probability 1, which is much better than any streaming algorithm can offer.
However, under the assumption that particles are not sorted in any way, the naïve solution would increase memory usage to terabytes even for the mesh with only $10^{12}$ cells in it.    

Finding frequent elements is one of the most studied problems in streaming settings, moreover, it is often used as a subroutine in many other algorithms \cite{liberty2013simple, ghashami2016frequent, indyk2005optimal, chakrabarti2010near, monemizadeh20101}. Let us first introduce the notation. The frequency (or count) of the element $i$ is the number of its occurrences in the stream $S$: $f_i = |\{j | s_j = i\}|$.
We will call element $i$ as $(\alpha, \ell_p)$-heavy if $f_i > \alpha\ell_p$ where $\ell_p = \left( \sum_j{f_j^p}\right)^{1/p}$. An approximate scheme for the problem is the following: 

\begin{problem}[Heavy Hitter]
Given a stream $S$ of $m$ elements the $\eps$-approximate $(\alpha, \ell_p)$-heavy hitter problem is to find a set of elements $T$, such that:
\begin{itemize}
\item $\forall i \in [n], f_i > \alpha \ell_p \rightarrow i\in T$ 
\item $\forall i \in [n], f_i < (\alpha - \eps) \ell_p \rightarrow i\notin T$ 
\end{itemize}
\end{problem}

Note that $\varepsilon$ in the definition above serves as slack for the algorithm to output some items which are not $(\alpha, \varepsilon)$-heavy hitters, but are ''$\varepsilon$ close'' to them. Typically smaller input $\varepsilon$ would cause the algorithm to use more memory.  
Finding the $k$ most frequent items in the stream is the same as finding all $(\alpha_k, \ell_1)$-heavy hitters, where $\alpha_k$ is the heaviness of the $k$-th most frequent item. Note that being $\ell_2$-heavy is a weaker requirement than being $\ell_1$-heavy: every $\ell_1$-heavy item is $\ell_2$-heavy, but the other way around it is not always the case. For example, consider the stream where all $n$ items of the dictionary appear only once. To be found in such a stream, a $\ell_1$-heavy hitter needs to appear more than $\varepsilon n$ times for some constant $\varepsilon$, while an $\ell_2$-heavy hitter needs to appear just $\varepsilon \sqrt{n}$. Catching an item that appears in the stream significantly less often is more difficult, thus finding all $\ell_2$-heavy hitters is more challenging than all $\ell_1$. 

The problem of finding heavy-hitters is well studied and there are memory optimal algorithms for $\ell_1$ \cite{misra1982finding,monemizadeh20101} and $\ell_2$\cite{chakrabarti2010near} heavy hitters, of which 
we are most interested in the latter. Here we will describe a Count Sketch algorithm \cite{chakrabarti2010near} which finds $(2\varepsilon, \ell_2)$-heavy hitters  $O(1/\varepsilon^2 \log^2(mn))$ bits of memory. 

\begin{figure}[t]
            \includegraphics[width=8cm]{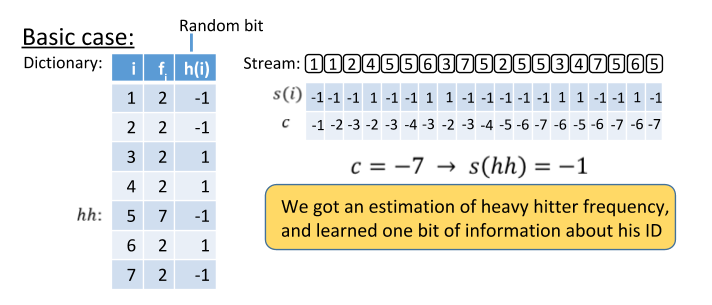}
            \centering
            \caption{Count Sketch subroutine on an example stream: each non-heavy item appears twice, heavy hitter (5) appears 7 times, a random $+1/-1$ bit is assigned to each item, the algorithm maintains the sum of the random bits, and the final sum is an unbiased estimator of the heavy hitter frequency having the same sign as its random bit}.
            \label{fig:CSintuition}
\end{figure}

\subsection{Count-Sketch Algorithms}
\begin{figure}[h]
            \includegraphics[width=7cm]{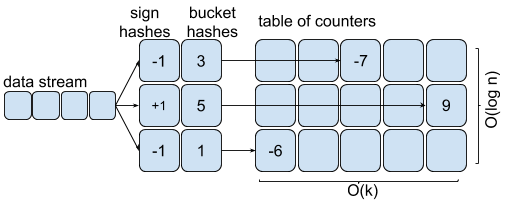}
            \centering
            \caption{Count Sketch algorithm scheme: bucket hash to identify the counter to which we should add the sign hash. Repeat $t$ times to recover the IDs.}
            \label{fig:CSscheme2}
\end{figure}

Consider a simplified stream with only one heavy item $i'$, and every other item $i$ appears in the stream only once. Let $h: [n]\Rightarrow \{-1, +1\}$ be a hash function which flips a $+1$/$-1$ coin for every item in the dictionary $i \in [n]$. If we will go through the stream $S= \{s_1, \ldots, s_m\}$ and maintain a counter $c = c + h(s_j)$, then at the end of the stream, with high probability, $c$ will be equal to the contribution of $i'$: $h(i')f_{i'}$ plus some small noise, while the majority of non-heavy contributors will be canceled by each other. The absolute value of $c$ can be considered as an approximation of the heavy item's frequency. At the same time, the sign of $c$ coincides with the random bit assigned to heavy items; thus, it helps us to reveal the ID of the heavy hitter by eliminating from consideration all items of the opposite sign. Simply repeating the experiment $t = O(\log n)$ times in parallel will reveal the entire ID of the heavy item. However, if the number of repetitions would be significantly smaller, we will face the problem of collisions, i.e. there will be items with the same random bits as the heavy item in all experiments. Thus we end up with many false positives due to the indistinguishability of those items from the heavy hitter. An example stream is depicted on Figure \ref{fig:CSintuition}. If our stream has $k$ heavy hitters, all we need to do is randomly distribute all items of the dictionary into $b = O(k)$ different substreams. Then with high probability none of the $O(k)$ substreams will have more than one heavy hitter in it, thus for each stream we can apply the same technique as before. On figure \ref{fig:CSscheme2} you can see the high-level intuition of both ideas described above: 
\begin{enumerate}
    \item Bucket hash to distribute items among $b$ substreams (buckets)
    \item Sign hash to assign random bit to every update
    \item Row of $b$ counters to maintain the sum of random bits
    \item $t$ instances to recover the IDs.
\end{enumerate}

Thus for each item update we need to calculate $t$ bucket hashes (specifying which substream/bucket this item belongs to) and $t$ sign hashes. We then update one counter in each row of the Count Sketch table, which is $t$ counters. In total the algorithm requires $t\cdot b = O(\log n)$ counters, which in turn requires $O(\log^2 n)$ bits of memory. For simplicity, here and later, we assume that $\log m = O(\log n)$, i.e. in our application mesh size is at most polynomially larger than the number of particles in the simulation). 

All of the statements above can be proven formally \cite{charikar2002finding}. Here we only show that using such a counter $c$ provides us with an unbiased estimator $\hat f_i = c\cdot h(i)$ for the frequency of the item~$i$: 
$$\forall i: E(c\cdot h(i)) = E\left(\sum_j{f_j\cdot h(j) h(i)}\right) = \sum_{i\neq j}{E(f_j\cdot h(j)h(i))} + f_i = f_i,$$ 
where the last equality is due to the 2-independence of hashing $h(\cdot)$. However, the variance of such estimators might be quite large and depends mainly on the second frequency moment of the other items in the substream. At the same time we know that with high probability  there is only one heavy hitter in each substream and we repeat the experiment $t = O(\log n)$ times. We take the median of those estimates, which reduces the variance and boost the probability for the final estimator to be within the approximation error from the real value. Summarizing, we have a data structure containing $b\times t = O(k)\times O(\log n)$ counters which maintain good estimates for the frequencies of the top $k$ most frequent items, but we still have to find the values of their IDs. There are three approaches to do this:
\begin{enumerate}
	\item \textbf{Count Sketch with Full Search(CSFS)} \\
		When all stream updates are processed we estimate the frequency of each possible item in the dictionary $i\in[n]$ and find the top $k$ most frequent.\\
			\textbf{pros}: updates are fast and easy to run in parallel \\
			\textbf{cons}: post-processing becomes very slow as the size of the dictionary grows 
	\item \textbf{Count Sketch with Heap(CSHe)} \\
		While processing each item, estimate its frequency and maintain the heap with the top $k$ most frequent items.\\
			 \textbf{pros}: post-processing takes zero time\\
			 \textbf{cons}: updates require extra log $k$ time-steps to update the heap
	\item \textbf{Count Sketch Hierarchical(CSHi)}\\
		Maintain two sketches, the first one for stream of super-items $S' = \{s_j/1000\}$ and the second one for the initial stream $S = \{s_j\}$. When all stream updates are processed, we first estimate the frequency of each possible super-item $i\in[n/1000]$ in the dictionary of $S'$ and find the top $k$ most frequent super-items $K' = \{hh'_j\}_{j=1}^k$, then estimate the frequencies of all potentially heavy items $i\in [n]\;\mbox{s.t.}\; i/1000\in K'$ and find the top $k$ most frequent items. This way we reduce the number of potentially heavy items to check. If necessary, more than 2 layers might be created.\\
		     \textbf{pros:} post-processing is fast even for very large dictionaries\\
			 \textbf{cons}: update time is $\rho$ times slower and the algorithm uses $\rho$ times more memory, where $\rho$ is the number of layers.  
\end{enumerate}  
CSFS contains set of $b\times t$ counters $M$, $t$ hash functions $h_s: [n]\rightarrow \pm 1$ and t hash functions $h_b: [n] \rightarrow [b]$ which decides which counter in the $t$-th row element $i$ corresponds to. In addition, CSHe contains the heap of pairs (item,frequency), and CSHi contains more than one sets of counters $\{M_i\}_{i=1}^\rho$. Let's define three following operations: 
\begin{itemize}
    \item Add($M$, $s_j$): \\ \tab $\forall i \in [t]: M_{i,h_{i,b}(s_j)} += h_{i,s}(s_j)$ 
    \item Estimate($M$, $j$):\\ \tab return $median\left(\left\{M_{i,h_{i,b}(j)}\cdot h_{i,s}(j)\right\}_{i=1}^t\right)$ 
    \item UpdateHeap($H$, $j$, $\hat f_j$): \\ 
    \tab if ($j \in H):\;\;\; H[j] := \hat f_j$ \\ \tab else if ($H$.top().$\hat f < \hat f_j$) : $\;\;H$.pop(); $H$.push($j, \hat f_j$);
\end{itemize}

The Add() operation updates all the counters,  Estimate() outputs current approximation for the frequency of the element $j$ and UpdateHeap() maintains the top $k$ most frequent items via updates of $(i, \hat f_i)$. The pseudo code for discussed functions is the following: 
\begin{algorithm}[h!]
    \caption{Count Sketch with Full Search(CSFS) }
        \begin{algorithmic}[1]
            \Procedure{Initialization}{}
                \State initialize $b\times t$ matrix of counters $M$ with zeroes
                \EndProcedure
            \Procedure{Processing the stream}{}
                \For {$s_i \in [m] = \{1, \ldots, m\}$} 
                    \State Add($M$,$s_i$)
                \EndFor    
            \EndProcedure
            \Procedure{Querying the data structure}{}
                \State initialize a heap $H$ of size $k$
                \For {$j \in [n]$} 
                    \State $\hat f_j$ = Estimate($M$, $j$);
                    \State UpdateHeap($H$ ,$j$ ,$\hat f_j$)
                \EndFor
                \For {$i \in [k]$} 
                    \State $(j, \hat f_j)$ = $H$.pop()
                    \State \Return $(j, \hat f_j)$
                \EndFor
            \EndProcedure

        \end{algorithmic}
\end{algorithm}
\begin{algorithm}[h!]
    \caption{Count Sketch with Heap(CSHe) }
        \begin{algorithmic}[1]
            \Procedure{Initialization}{}
                \State initialize $b\times t$ matrix of counters $M$ with zeroes
                \State initialize a heap $H$ of size $k$
                \EndProcedure                        
            \Procedure{Processing the stream}{}
                \For {$s_i \in [m] = \{1, \ldots, m\}$} 
                    \State Add($M$,$s_i$)
                    \State $\hat f_j$ = Estimate($M$, $s_j$)
                    \State UpdateHeap($H$ ,$s_j$ ,$\hat f_j$)
                \EndFor    
            \EndProcedure
            \Procedure{Querying the data structure}{}
                \For {$i \in [k]$} 
                    \State $(j, \hat  f_j)$ = $H$.pop()
                    \State \Return $(j, \hat f_j)$
                \EndFor    
            \EndProcedure
        \end{algorithmic}
\end{algorithm}
\begin{algorithm}[h!]
    \caption{Count Sketch Hierarchical(CSHi) }
        \begin{algorithmic}[1]
            \Procedure{Initialization}{}
                \State initialize two $b\times t$ matrices of counters $M_1$ and $M_2$ with zeroes
                \EndProcedure
            
            \Procedure{Processing the stream}{}
                \For {$s_i \in [m] = \{1, \ldots, m\}$} 
                    \State Add($M_1$,$s_i/1000$)
                    \State Add($M_2$,$s_i$)
                \EndFor    
            \EndProcedure
        
            \Procedure{Querying the data structure}{}
                \For {$j \in [n/1000]$} 
                    \State $\hat f_j$ = Estimate($M_1$, $j$);
                    \If{$\hat f_j > \theta_1$}
                        \For {$j' \in [1000j:  1000(j+1)]$} 
                            \State $\hat f_{j'}$ = Estimate($M_2$, $j'$);
                            \If{$\hat f_{j'} > \theta_2$}
                                \State \Return $(j', f_{j'})$
                            \EndIf                    
                        \EndFor
                    \EndIf
                \EndFor    
            \EndProcedure
        \end{algorithmic}
\end{algorithm}

Similar construction is used in the algorithm Count Min Sketch \cite{cormode2005improved}. The algorithm utilize the similar logic and the same size table of counters, however for each update it computes only one hash (to specify the bucket to be updated) rather than two in Count Sketch, and output the minimum over the estimates, rather than the median. Thus subroutines ''Add'' and ''Estimate'' are different:
\begin{itemize}
    \item Add($M$, $s_j$): \\ \tab $\forall i \in [t]: M_{i,h_{i,b}(s_j)} += 1$ 
    \item Estimate($M$, $j$):\\ \tab return $\min\left(\left\{M_{i,h_{i,b}(j)}\cdot h_{i,s}(j)\right\}_{i=1}^t\right)$ 
\end{itemize}
We compare Count Sketch and Count Min Sketch experimentally. However the latter only finds $\ell_1$ heavy hitters, so we expect it to be outperformed by Count Sketch. 

\subsection{Choosing the parameters for the algorithms}
To calculate the parameters for the algorithm resulting in a certain desired number of heavy hitters we need to make an estimate of the density distribution of counts in cells. 
Cosmological simulations begin with an almost uniform lattice of particles. As time goes on, gravity pulls particles toward tiny density fluctuations of the early universe.
As for example shown in \cite{neyrinck2009logn}, the resulting density distribution can be well modelled with a log-normal PDF.
\[
P_{\rm LN}(\delta) =\frac{1}{(2\pi\sigma_1^{2})^{1/2}}\cdot\exp{\frac{-[\ln(1+\delta) + \sigma_1^2/2]^2}{2\sigma_1^2}}\cdot\frac{1}{1+\delta},
\]
where $\delta = \rho/\bar{\rho}-1$ is the over-density ($\bar{\rho}$ is the average density), $\sigma_1^2(R)$ is the log-transformed variance of density in a sphere of radius $R$. This formula can be used as a qualitative indication of the particle distribution on a cubic grid as well, and was also assumed in \cite{liu2015streaming}. 
Based on this model, we compute the minimum number of points in a ``halo cell" (e.g., with over-density $\geq 200$, the value corresponding to halos that have just virialized). We denote this as $N$. 
Ideally we would then compute the square 2-norm of all counts in cells, i.e., $Z = \sum_{c: \text{ as a cell}}$(number of points in $c$)$^2$. However the claim of this paper is that this may not be possible to do exactly, hence we assume
that this value can also be predicted by the log-normal distribution.
Next, we compute $\alpha = N^2 / Z$. This is the so-called heaviness, i.e., the counts in a halo-cell is at least $\alpha$ fraction of the sum-of-square of the total counts. It can be used to determine the $t$ and $b$ parameters used later in this paper
(in particular we shall set $b\approx 1/\alpha, t\approx\log N$). We shall set the target number of heavy hitters we want to detect as $k\sim 1/\alpha$, i.e., the top-$k$ heavy cells contain the heavy hitters.
By using standard cosmological parameters for $\sigma_1$ (see \cite{liu2015streaming} for details), we can calculate that for cell size of $\approx 1$Mpc/h, $\alpha \approx 1/1000$.
Note that we cannot determine the exact constants for $t$ and $b$ since they are algorithm-dependent. We thus leave them as tunable parameters. 
 
\section{Implementation}
\label{sec:implementation}
Paper~\cite{liu2015streaming} presented a halo finding tool using streaming algorithms that can be very useful even in systems with as low as 1GB memory. However, the running time of that tool was more than 8 hours on a desktop for a relatively small dataset. Here we provide a new algorithm based on an efficient GPU implementation. The core part of the halo finding tool relies on the implementation of the Count Sketch algorithm. All experiments in this section were carried out on the CPU Intel Xeon X5650  @ 2.67GHz with 48 GB RAM and GPU Tesla C2050/C2070.

\subsection{Count Sketch Implementation}
The data flow of the Count Sketch algorithm consists of 5 basic stages, i.e. for each item we need to do the following:
\begin{enumerate}
    \item Compute cell ID from XYZ representation of the particle
    \item Compute $t$ bucket hashes and $t$ sign hashes 
	\item Update $t$ counters
	\item Estimate the current frequency for the item (find median of $t$ updated counters)
	\item Update the heap with current top-$k$ if necessary
\end{enumerate}
Below we consider different implementations of the Count Sketch algorithm with the argument to architectural decisions made:
\begin{enumerate}
	\item CPU: \\
		Purely CPU version of the Count Sketch has all five stages implemented on the CPU and described in details in \cite{liu2015streaming}. As depicted below, it takes 8.7 hours to process one snapshot of all particles from the Millennium dataset. In the breakdown of the profiler output below, where integer numbers denote the 5 stages of the Count Sketch algorithm and fractions show proportional amounts of time spent on that stage, we can see that the second stage is computationally the most expensive. The most straightforward improvement is to ''outsource'' this computation to the GPU. We implemented this idea, and we describe it further below.
		\begin{figure}[h]
            \includegraphics[width=5cm]{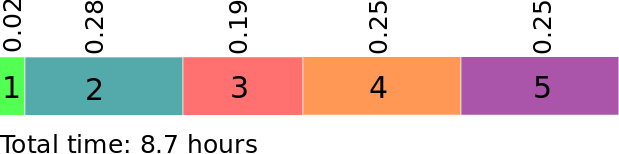}
            \centering
        \end{figure}
        
	\item CPU + hashes on GPU \\
		 In this implementation, we are trying to ``outsource'' the most time intensive operation --- calculating hashes. Recall that we need to compute $2t$ hashes. As long as $t$ is a relatively small number ( $\le16$), a naive parallelism which suggests computing all hashes for each particle in $t$ parallel threads, will not provide a significant speed up due to the inability to saturate all cores ($\sim2000$) of the graphics card. Thus to improve performance even further, we need to make use of data parallelism, which assumes computing hashes for a batch of updates at the same time. Such an approach is straightforward due to the fact that computing hashes are identical operations required for all particles and those operations can be performed independently. As illustrated below, the GPU computes hashes almost for free, compared to stages 3,4 and 5, and total time drops by $35\%$. The next bottleneck is stage 3, during which the algorithm updates counters. Although it is just $2t$ increments or decrements, they happen at random places in the table of counters. This makes it impossible to use CPU cache and memory access becomes a bottleneck for the algorithm.
		 \begin{figure}[h]
            \includegraphics[width=5cm]{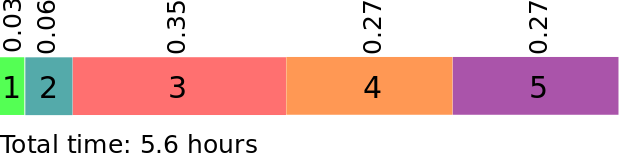}
            \centering
        \end{figure}

	\item GPU + heap on CPU \\
		Updating counters (stage 3) and estimating current frequencies (stage 4) are two very connected stages. If we keep them together we can significantly save on the number of queries to the memory. Implementing a time efficient heap (stage 5) on the GPU is quite challenging, due to hardware features. Thus our next implementation takes advantage of the CPU for maintaining the heap, while doing all other computations and storing the table of counters on the GPU. The basic data flow can be described as follows:
		\begin{enumerate}
            \item CPU sends a batch of particles in XYZ representation onto GPU
            \item GPU processes all particles in parallel: compute cell ID, compute hashes, update counters and estimate frequencies
            \item GPU sends a batch of estimates back to the CPU
            \item CPU maintains heap with top $k$ items using estimations from GPU
	    \end{enumerate}
		It can be seen below that adopting such an approach pushed the total time of the algorithm down to 38 minutes. In the breakdown of the profiler, one can see that updating the heap became a new bottleneck for the algorithm.
		\begin{figure}[h]
            \includegraphics[width=5cm]{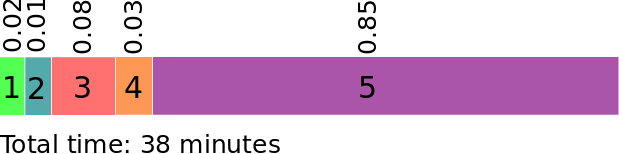}
            \centering
        \end{figure}
        
	\item GPU without heap  \\
		While heap on the CPU is quite efficient, it still slows down the process quite seriously, especially when the top $k$ gets larger and reaches $10^6$. On large datasets this might cause many items to have an update time close to $\log k$.  Moreover, keeping the heap on the CPU forces the GPU to send a lot of data back to the CPU. Avoiding this data transfer would improve the slowest memory operation by a factor of 2. Thus we decided to switch from Count Sketch with Heap (CSHe) to Count Sketch with Full Search (CSFS), both of which were broadly described in the previous section. The CSFS algorithm works in two modes: update mode, which encompasses calculating hashes and updating counters, and estimate mode, which deals with estimating the frequency for all cells and emitting the top $k$. The CSFS algorithm is first invoked in update mode for the entire stream, and when it finishes, the generated table of counters is used as input to estimate mode. While in estimate mode, we still need to maintain the top $k$ items and do it on the GPU. This can be done semi-dynamically by adding to the array all items which are larger than some threshold. Then, if we have more than $k$ items, we will raise the threshold and rearrange elements in the array, deleting those items which do not satisfy the new threshold. If we grow the threshold geometrically we can guarantee that such ''cleaning'' step won't happen too often. Such an approach cannot be applied to the CSHe algorithm due to the possibility of two updates for the same cell. In the figure below, the stream time, which includes only the update mode, takes only $3.5$ minutes, while the estimate mode takes $25$ minutes. The time of the estimate mode, i.e. query time, linearly depends on the size of the mesh, due to the necessity to estimate the frequency for every cell in the mesh. For example, in the same experiment for the mesh with $5\cdot10^8$ cells, query time would be less than $10$ seconds. 
		
		\begin{figure}[h]
            \includegraphics[width=5cm]{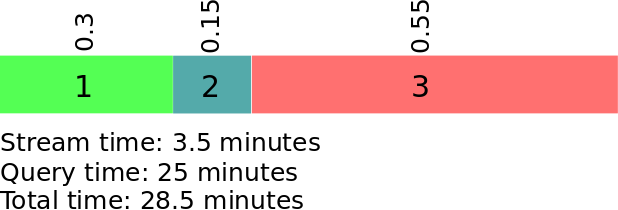}
            \centering
        \end{figure}
        
	\item GPU hierarchy \\
    	 As it was already discussed in the previous section, one of the ways to decrease query time is to eliminate the full search and implement it as a search tree instead. In our case, the search tree (hierarchy) will contain only two layers. By grouping cells together we can find the heavy super-cells first (using a small mesh), then search for heavy cells only inside heavy super-cells. We will merge cells by their IDs in the top layer with a dictionary size of $\sim 10^8$, find top $c\cdot k$ super-cells and find top $k$ cells inside the selected heavy super-cells, where $c>1$ is a small constant. As can be seen below, such an approach reduces query time from 25 minutes down to 55 seconds. However, it requires twice the amount of memory due to the need to store a table of counters for each layer. It can also be observed that time performance of the update mode gets worse, due to the necessity to calculate twice as many hashes and update twice as many counters. The total time of the algorithm is 5 minutes, which is very impressive for the size of the dataset and the mesh.
    	 The total performance improvement over the sequential CPU implementation is more than $100$-fold.
    	\begin{figure}[h]
            \includegraphics[width=5cm]{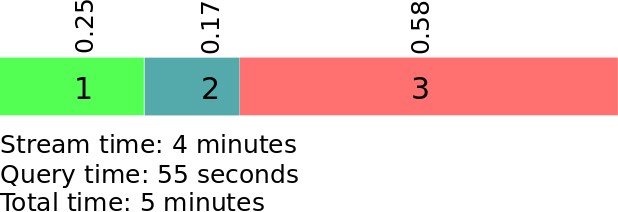}
            \centering
        \end{figure}
		
\end{enumerate}

Here we will briefly introduce the key architectural decisions in the implementation of the ``GPU without heap'' version of the algorithm. While it is not the most efficient implementation, it is easier to explain. At the same time, it makes it straightforward how to extend it to the "hierarchical" version. The graphical processor is a separate device that has many limiting features compared to the CPU. In this project, all our experiments leverage the CUDA platform to make use of the graphical processor's capabilities. \cite{nvidia2010programming}. 

A GeForce GTX 1080 has 20 multiprocessors (SM) each with 128 cores (threads). CUDA introduced a block/thread approach, such that all computations are grouped into blocks, where one block is always implemented on only one SM. Within a block, we can specify how to share computation between threads. CUDA has three layers of memory: 
\begin{enumerate}
    \item \textbf{Global memory:}
        accessible within the device and conventionally is quite large (up to $8$ GB). It is also the only type of memory that can be used to copy to or from RAM. At the same time, it is the slowest memory on the device.
    
    \item \textbf{Shared memory:}
        accessible from within the block and shared among all threads of that block. Shared memory is $\sim 10$ times faster than global memory, however, it is very limited with $\sim 48-64$KB per SM. 
 
    \item \textbf{Registers:}   
        there are $2^{15}$ 32-bit registers per SM. They are as fast as shared memory, but visible only to the thread.  
\end{enumerate}

Storing a table of counters for Count Sketch is possible only in global memory. Primarily, this is due to the large size of the counters $\sim 1$GB. Secondly, counters are accessed in random order, which makes it impossible to store some localities in the shared memory. In our implementation, each block is in charge of exactly one update of the stream. In order to make an update, one needs to calculate $2t$ hash functions and update $t$ counters, thus we distributed this work among $t$ threads, each calculating two hashes and updating one counter. 

Note that to avoid memory access conflicts we need to use atomic operations, which are present in CUDA. However we expect the number of conflicts not to be very large: while the typical width of the table is $10^7$ counters and the maximum number requests is bounded by the number of GPU threads (which is $\sim2000$ in our case), the probability of collision is negligible. In practice, we can see that using non-atomic operations would give us at most a $10\%$-fold gain in time performance. The pseudo code for each thread is presented in Algorithm \ref{algo:gpuCode}.

After the stream is processed, we need to find the IDs of the heavy hitters. As described earlier, we need to find an estimation for each item in the dictionary. Here we will use the same approach as for stream processing. Each block will be in charge of one cell. Each thread will be in charge of an estimation based on one row of Count Sketch counters. The procedure for each thread is described in Algorithm \ref{algo:gpuCode}.

\begin{algorithm}
    \caption{GPU thread code for Count Sketch}
        \begin{algorithmic}[1]
            \Procedure{Update}{cellID}
                \State $i =$ threadID;
                \State $M[i, h_{i,b}(cellID)] += h_{i,s}(cellID)$
            \EndProcedure
            \State{}
            \Procedure{Estimate}{cellID}
                \State \textbf{shared} estimates[t];
                \State \textbf{shared} median;
                \State $j =$ threadID;
                \State $\hat f = M[i, h_{j,b}(cellID)]\cdot h_{j,s}(cellID)$;
                \State estimates$[j] = \hat f$;  
                \State \textbf{synchronize}
                \State \textbf{int} above, below $= 0$;  
                \For {$i \in [t]$} 
                    \State below $+= ($estimates$[i] < \hat f)$
                    \State above $+= ($estimates$[i] > \hat f)$
                \EndFor    
                \If{above $<= t/2$ and below $<= t/2$} 
                    \State median $= \hat f$ 
                \EndIf
                \State \textbf{synchronize}
                \If{$j = 1$ and median $> \theta$} 
                    \State \Return median
                \EndIf
            \EndProcedure
    \end{algorithmic}
    \label{algo:gpuCode}
\end{algorithm}

Note that we find the median using a very naive algorithm --- for each item of the array check if it is a median by definition. That is thread $i$ would be in charge of checking if the number of estimates smaller than estimates$[i]$ is equal to the number of estimates larger than estimates$[i]$, and reporting/recording the found media if so.
This is one of the reasons why all estimates should be reachable by all threads, and thus should be stored in the shared memory. While in sequential implementation this approach would take $O(t^2)$ time steps, here we use $t$ parallel threads, ending up with time complexity of $O(t)$.

\begin{figure}[h]
            \includegraphics[width=8cm]{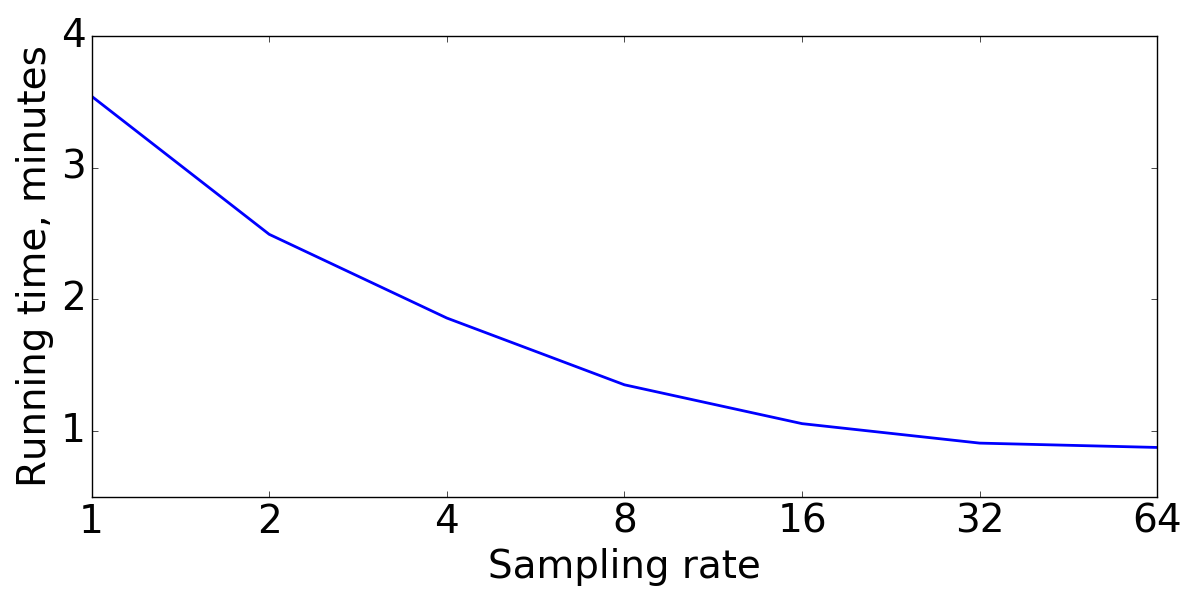}
            \centering
            \caption{Dependency of time performance on sampling rate.}
            \label{fig:samplingVStime}
\end{figure}

To boost the time performance even further we can apply sampling. However, one should not expect performance to improve linearly with the sampling rate, because of necessity to compute sampling hashes for all particles. The dependency of the time performance on the sampling rate is depicted in figure \ref{fig:samplingVStime}. From that graph, one can see that changing the subsampling rate from 8 to 16 is the last significant improvement in time performance.

\section{Evaluation}
\label{sec:evaluation}
In this paper, we present a tool which is capable of finding up to  $10^5-10^6$ densest cells in state of the art cosmological simulations for an arbitrary sized regular mesh. Moreover, the proposed technique makes these procedures available even for the desktop or a small server. In this section, we evaluate this claim. We do this in two steps. In the first, which we call the {\em algorithmic evaluation} we compare the rank order produced by the heavy hitter algorithm directly to the exact results. In the second, we perform a scientific evaluation and analyze what the effects are of the randomized nature of the approximate algorithm to various statistical measures of astrophysical interest; namely the tail end of the counts-in-cell distribution and the spatial clustering of excursion sets. 

\subsection{Evaluation Setup}
For testing and evaluation, we use the Millennium dataset \cite{lemson2006halo} with $10^{10}$ particles in a cube with side length $500$ \mpch. The cell size in the grid is 0.1 \mpch, thus the total grid contains $1.25 \times 10^{11}$ cells. Our goal is to find top $10^5$ to $10^6$ heaviest cells. Those numbers are important to understand some decisions in choosing the specific architecture of the implementation.

The data, originally stored in the GADGET\cite{springel2005cosmological} format, is reorganized, such that every 64 bits contains 3 coordinates for one particle. This reorganization helps to reduce the number of global memory writes inside the GPU. After such a reorganization the entire dataset weights in  at 80 GB. One of the time performance bottlenecks in such settings is reading data from the hard drive. We implemented a parallel I/O system that includes 3 SSDs and 24 HDDs without data replication, and this way we reduced the pure I/O time from 15 minutes to 20 seconds. 
For comparison purposes all experiments were accomplished on two different hardware configurations:
\begin{enumerate}
\item AMD Phenom II X4 965 @ 3.4 GHz, 16 GB RAM, GPU GeForce GTX 1080.
\item Intel Xeon X5650  @ 2.67GHz, 48 GB RAM, GPU Tesla C2050/C2070.
\end{enumerate}

\subsection{Top-k Cells}
First, let's introduce different ways of finding the top $k$ most frequent cells with exact counts. Given a set of particles in the simulation box and a regular grid of a fixed size we need to find $k$ cells of the grid containing the largest numbers of particles, together with the IDs of those cells.  The algorithm is required to return an estimate of the number of particles in each cell. The most straightforward solution to this problem is to count the number of particles in each cell precisely. Such an exact algorithm might be described as follows: 
\begin{enumerate}
    \item Create a counter for every cell in the grid
    \item While making a pass through the dataset, update the cell counters based on the position of the particles
    \item Find $k$ ''heaviest'' cells and return their IDs and exact counts
\end{enumerate}
This solution breaks down in step~1 once the size of the mesh is too large to store all counters in memory. 

It is possible to remove the memory problem at the expense of worsening time performance by making multiple passes over the data, as in the following algorithm. Assuming the memory is about a factor $1/\lambda$ of the total size of the grid: 
\begin{enumerate}
    \item Create a counter for every cell in the range $[i-n/\lambda, i]$, and use the basic algorithm above to find $k$ ''heaviest'' cells in the range and call them $topK_i$.
    \item Repeat previous step for all ranges $i\in\{n/\lambda, 2n/\lambda, \ldots, n\}$ and find top $k$ ''heaviest'' cells in $\cup_i topK_i$
\end{enumerate}



This multi-pass trick becomes unfeasible when the size of the mesh grows too large compared to the available memory, as it would take too many passes over the data. However, to evaluate how well our algorithm approximates the exact top $k$ with counts, we do need to have exact counts. That was one of the reasons why  \cite{liu2015streaming} restricted themselves to relatively small meshes. In the current paper, we show results for meshes of sizes $10^8$,  $10^{11}$ and $10^{12}$. We will provide algorithmic evaluation only for $10^8$, where the naive precise algorithm can be applied, and for $10^{11}$, where we apply the trick described above and do 20 passes over the dataset. For the mesh of size $10^{12}$, algorithmic evaluation is more challenging and therefore only the scientific evaluation will be performed. 
\subsection{Evaluation of algorithm:}

\begin{figure}[h]
\includegraphics[width=8cm]{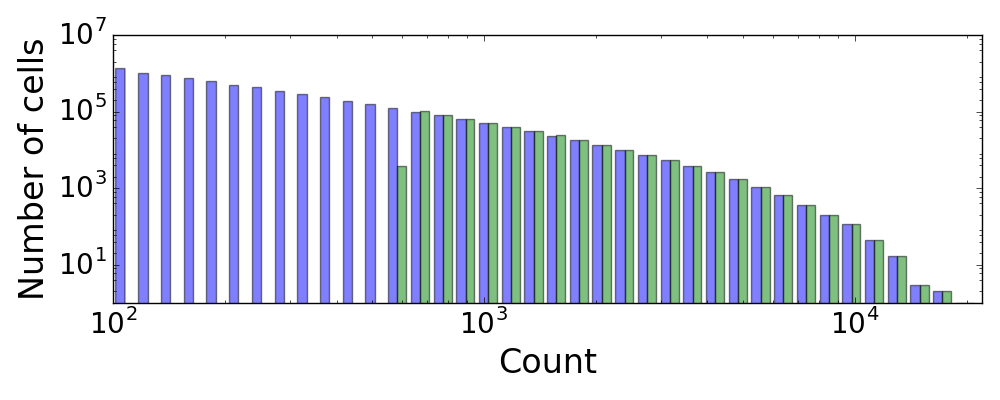}
\centering
\caption{Cell density distribution for the top $0.5 \cdot 10^6$ cells found by Count Sketch (in green) and the top $10^7$ cells found by exact counting (in blue). }
\label{fig:cellsHist}
\end{figure}

In this section, all experiments are for the mesh size $10^{11}$. Fig.~\ref{fig:cellsHist} shows the distribution of exact cell counts for the top $10^7$ cells.

\begin{figure}[h]
\centering
  \begin{subfigure}[b]{0.4\textwidth}
  \includegraphics[width=1\linewidth]{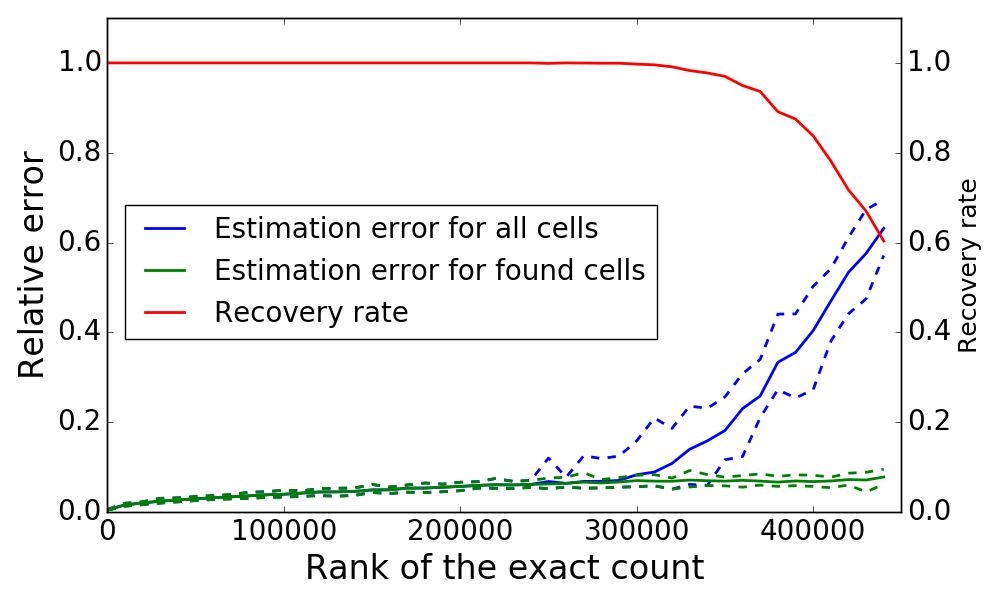}
  \caption{Cell size = 0.1 \mpch}
  \label{fig:relError} 
\end{subfigure}

\begin{subfigure}[b]{0.4\textwidth}
\centering
    \includegraphics[width=1\linewidth]{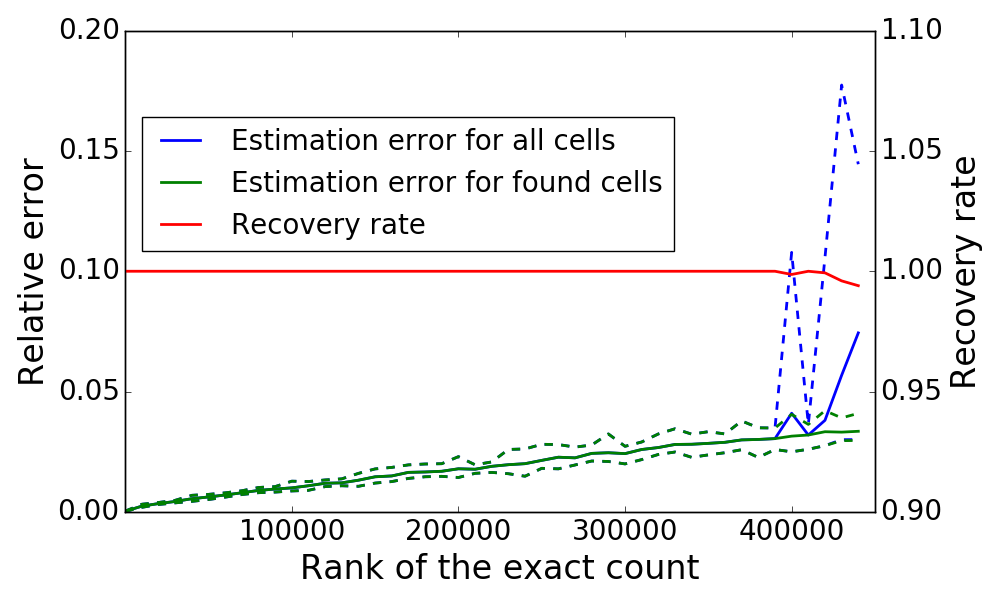}
    \caption{Cell size = 1 \mpch}
    \label{fig:relError500}
\end{subfigure}

\caption{Relative error vs. rank for (a) cell size 0.1\mpch~~and (b) cell size 1\mpch.~ Each experiment was carried $20$ times. Dashed lines depict the maximum and the minimum, while the solid line shows the average over those $20$ runs for each rank value.}
\end{figure}

\begin{figure}[t]
  \includegraphics[width=8cm]{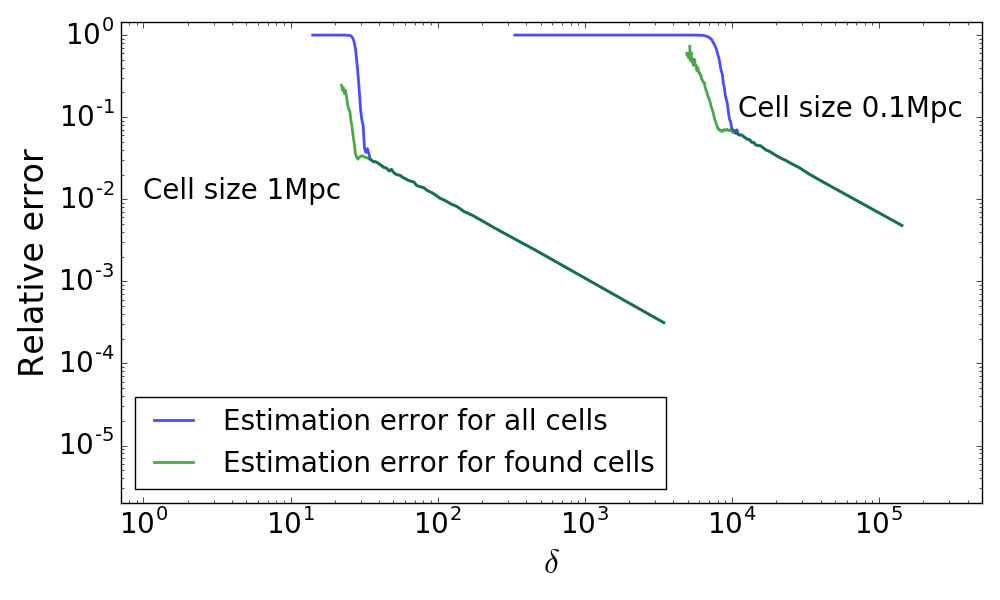}
  \centering
  \caption{Relative error vs. $\delta$ of the cell}
  \label{fig:relErrorDelta}
\end{figure}

\begin{figure}[t]
    \includegraphics[width=8cm]{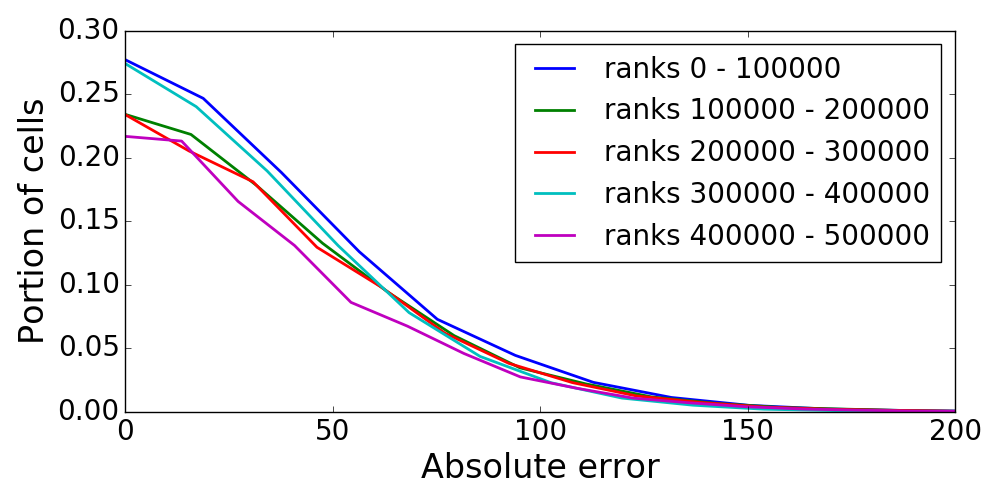}
    \centering
    \caption{Distribution of absolute error for different ranks}
    \label{fig:absErrorDistribution}
\end{figure}

Most experiments in this section use a Count Sketch with parameters $t = 5, b= 10^7$ and $k = 5\cdot10^5$. A motivation for these values will be provided later. To understand how well the Count Sketch approximates the exact counts and how well it reproduces the rank order, we determine how the relative error grows with the rank inside the top-$k$ cells. To do so, for each cell $i$ we find its count $c_i$ and rank $r_i$ in the output of an exact algorithm and its count $\hat c_i$ in the Count Sketch output. If cell $i$ is not present in the Count Sketch output, i.e. not among its top $k$ heaviest cells, we define $\hat c_i = 0$. Fig.~\ref{fig:relError} shows, in blue, the dependency on rank $r_i$ of the relative error, defined as $ |c_i - \hat c_i|/c_i$. Here we use a bin size of 100 in $i$ for the averaging. 

The relative error is shown in green and is determined for cells which were among the top $k$ for both the exact and the Count Sketch counts. By ignoring the cells not found in the Count Sketch, the relative error is artificially reduced. On the other hand, treating those cells as empty $\hat c_i = 0$ pushes the error rate up significantly.  This overestimates the error compared to the count that might have been determined had the Count Sketch included those cells, for example by using a larger value of $k$. 

As we can see, up to a rank of $\sim 250000$ the algorithm works reliably and has quite low approximation error. However, at higher ranks the error grows rapidly. The main cause of this is the loss of heavy cells, rather than a bad approximation of the counts for the cells that were accepted by the Count Sketch. This is shown by the fact that the green line remains low. 

Fig.~\ref{fig:relErrorDelta} shows the same graphs, but now plotted against the over-density $\delta_i = (N_i-<N>)/<N>$ in cells. This quantity is more meaningful from an astrophysical point of view compared to the rank. It shows that the errors are stable for a large range of over-densities, but very quickly shoot up near a threshold. That threshold depends on the size of the cell as the comparison in Fig.~\ref{fig:relErrorDelta} shows. Note that the size of the cell for any specific dataset would influence the number of particles in the each of the top-$k$ heavy cells and $\ell_2$ norm of the stream. 

There is a straightforward reason why the approximate algorithm loses so many heavy cells. Before we explain it we need point out three important facts. First, Fig.~\ref{fig:absErrorDistribution} shows that the absolute error is about constant for all cells. This can be understood from the theoretical arguments in \cite{chakrabarti2010near}, which state that all estimations have an additive approximation error of $\varepsilon \ell_2$, i.e. for each cell, error does not depend on the count, but only on the $\ell_2$ norm of the entire dataset. 
Second, as we can see from Fig.~\ref{fig:cellsHist}, the number of cells is increasing exponentially with the count going down (this is a property of the mass function in the cosmological simulation). Third, for the cells with ranks near 250000, the actual count is $\sim 820$, and for the cells with ranks near 500000 the actual count is $\sim 650$. While searching for the top-$k$ cells using Count Sketch estimations we will face two types of errors:
\begin{enumerate}
    \item[type 1:]false rejection of heavy cells caused by {\em underestimation} of the true count due to approximation error
    \item[type 2:]false exclusion of heavy cells caused by {\em overestimation} of counts of cells below the top-K selection criterion.
\end{enumerate}

\begin{figure}
\centering
  \begin{subfigure}[b]{0.4\textwidth}
  \includegraphics[width=1\linewidth]{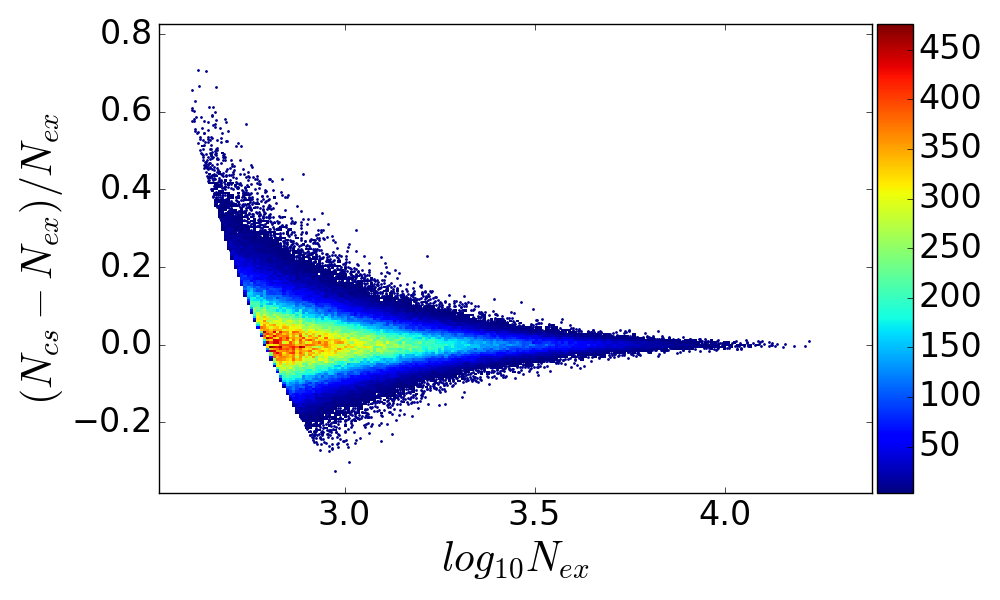}
  \caption{cell size = 0.1 \mpch}
  \label{fig:cloudComparison1} 
\end{subfigure}

\begin{subfigure}[b]{0.4\textwidth}
  \includegraphics[width=1\linewidth]{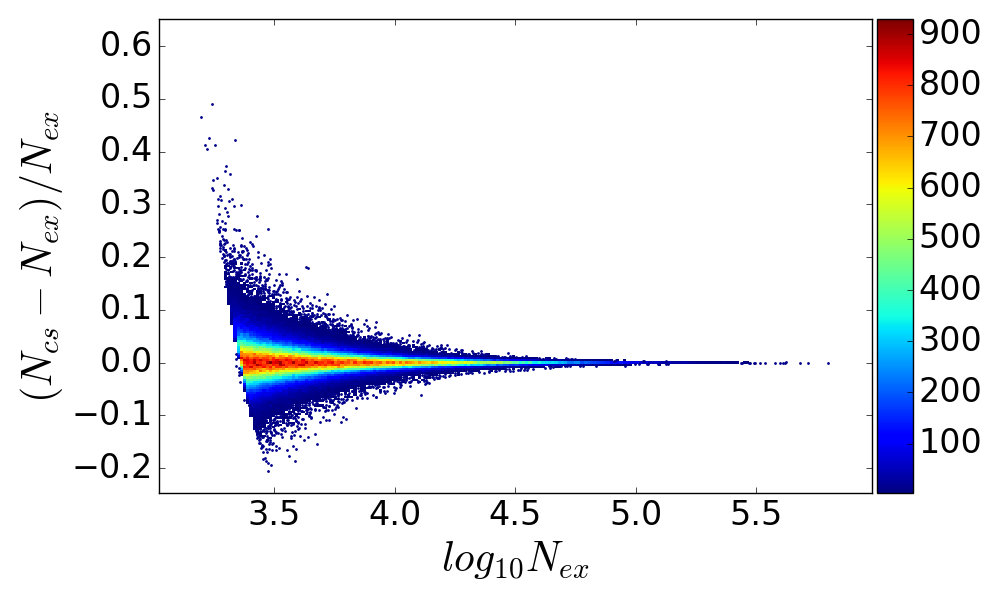}
  \caption{cell size = 1 \mpch}
  \label{fig:cloudComparison2}
\end{subfigure}

\caption{ Count distortion for the cell size = 0.1 \mpch~~on the top and for the cell size = 1 \mpch~~on the bottom.}
\end{figure}

We expect that having 250000 elements with counts between $\sim 650$ and $\sim 820$ with an average approximation error $\sim 80$ (and ranging up to 250) will cause significant loss of heavy cells in the top $k$. We can see this in Fig.~\ref{fig:relError}, which also depicts the recovery rate. Thus we can conclude that the main cause for missing heavy cells in the output is the fact that many cells have counts which are relatively close to the approximation error of the algorithm. We have tested this conclusion by running the algorithm for larger cell sizes, with significantly larger expected counts. This increases the typical $|c_i - c_j|$ for cells $i$ and $j$ with small rank distance. The result of an experiment with a cell size of 1 \mpch\ is shown in Fig.~\ref{fig:relError500}, which corroborates our hypothesis. The difference in the results for different mesh sizes is even more obvious in the relative error vs. exact count graphs in Fig.~\ref{fig:cloudComparison1} and Fig.~\ref{fig:cloudComparison2}. Note that closer to the cut-off threshold the algorithm is tending to overestimate the count rather then underestimate. This behaviour is reasonable due to the fact that only one-way error is passing the threshold test, while all items with underestimated counts are discarded by the algorithm.

\begin{figure}[h]
\includegraphics[width=8cm]{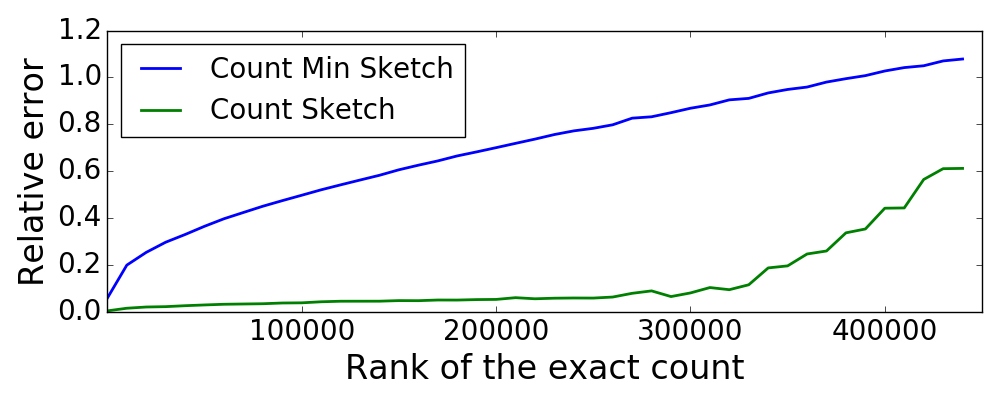}
\centering
\caption{Relative error for the counts in the output of the
Count Sketch algorithm and Count Min Sketch algorithm, cell size = 0.1 \mpch}
\label{fig:relErrorCMSvsCSH}
\end{figure}

We know that every $\ell_2$ heavy hitters algorithm catches all $\ell_1$ heavy items and some items which {\em are} $\ell_2$-heavy but not $\ell_1$-heavy. While asymptotically the space requirements for both algorithms are the same, the time performance for $\ell_1$ algorithms is better in practice than for the $\ell_2$ algorithms. It is therefore of interest to compare the two in the specific application to our problem. To do so we compared the Count Sketch with the intuitively similar $\ell_1$  Count Min Sketch algorithm. Fig.~\ref{fig:relErrorCMSvsCSH} shows that the approximation error differs significantly, with the Count Sketch algorithm giving much more accurate results.  

\begin{figure}[h]
\includegraphics[width=8cm]{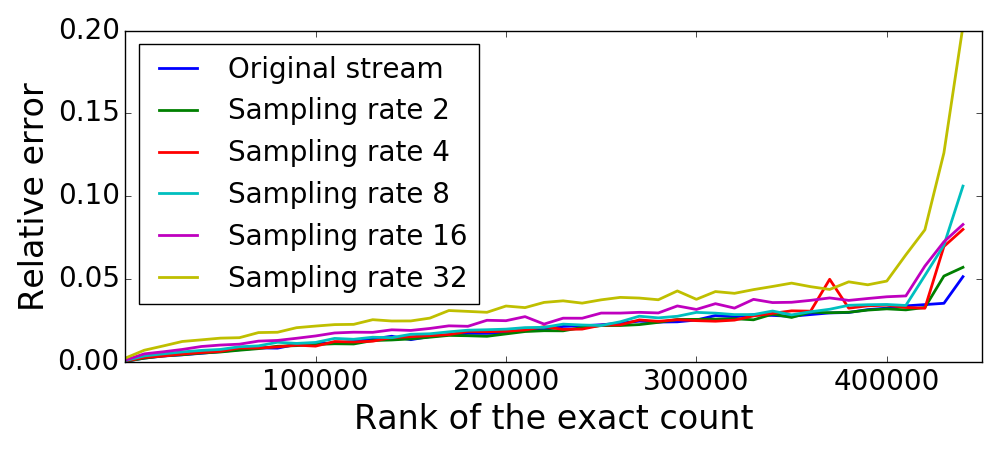}
\centering
\caption{Relative error for the counts in the output of 
the Count Sketch algorithm with different sampling rates, cell size = 1 \mpch}
\label{fig:samplingVSerror500}
\end{figure}

\begin{figure}[h]
\includegraphics[width=8cm]{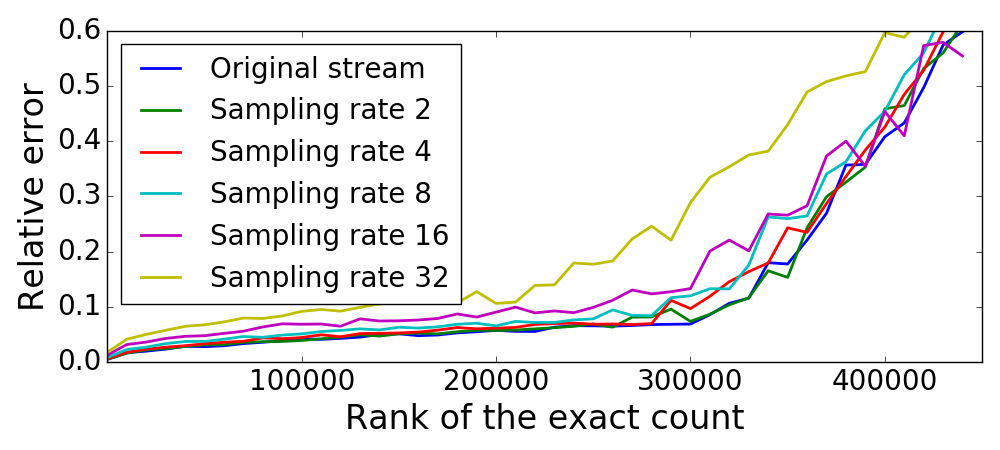}
\centering
\caption{Relative error for the counts in the output of 
the Count Sketch algorithm with different sampling rates, cell size = 0.1 \mpch}
\label{fig:samplingVSerror}
\end{figure}

As it was mentioned earlier, a random sampling of the particles before feeding them to the algorithm can significantly improve the time performance of the entire procedure. To investigate the influence of such sampling on the approximation error we carried out experiments comparing different sampling rates for mesh sizes 1 \mpch\ and 0.1 \mpch. From the figures \ref{fig:samplingVSerror500} and \ref{fig:samplingVSerror} we can see that in both cases a sampling rate of $1/16$ still provides a tolerable approximation error. It is important to recall that the time performance does not scale linearly with the sampling rate due to the need to compute the sampling hash function for each element. This operation is comparable in time to processing the element through the entire data flow without skipping the Count Sketch.

The crucial advantage of the algorithm presented here compared to existing algorithms is the improvement in memory usage. Traditional algorithms often require complete snapshots to be loaded into memory, which for state-of-the-art cosmological simulations implies they cannot be analyzed on a  small server or even one desktop. 
For a cell size of 1~\mpch$\ $ and a box size of ~500\mpch$\ $  our mesh would contain only $1.25 \cdot 10^8$ cells which require only $500$ MB for a naive algorithm and provides an exact solution. Such a low memory footprint makes the naive solution feasible even for a laptop. For a cell size of 0.1 \mpch with the same box size the memory requirements would be $1000\times$ larger and barely fit onto a mid-size server. 

\begin{figure}[h]
\includegraphics[width=8cm]{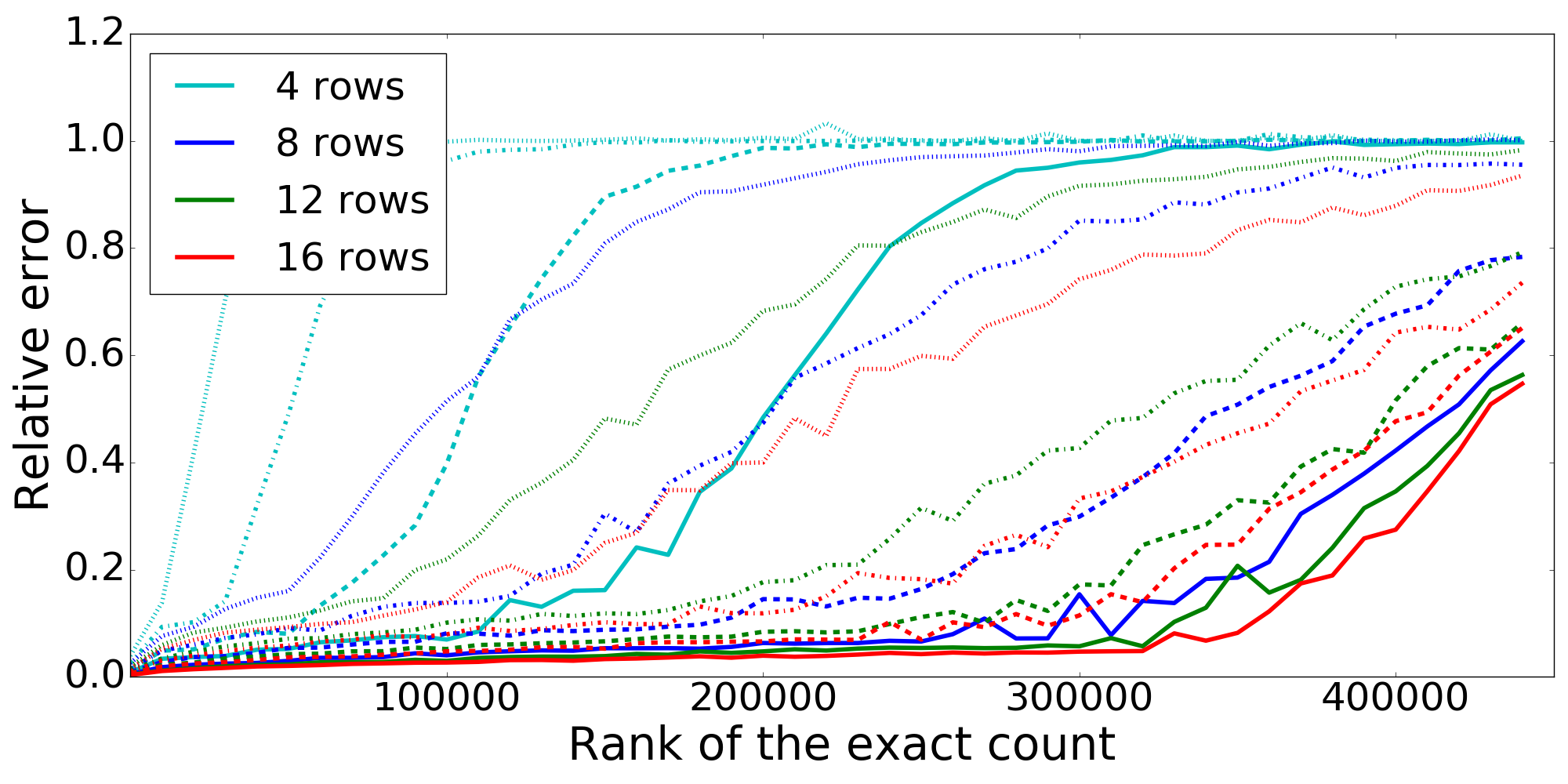}
\centering
\caption{Relative error for the counts in the output of the
Count Sketch algorithm with different internal parameters, cell size = 0.1Mpc. Color is the height of the CS table, and line type is the width of CS table: solid is $16\cdot 10^6$, dashed is $8\cdot 10^6$, dash-dotted is $4\cdot 10^6$,  and dotted is $10^6$ columns}
\label{fig:relErrorVsCSparams}
\end{figure}

Next, we investigate the trade-off between time performance, memory requirement and approximation error in more detail. 
The Count Sketch data structure consists of $t\times b$ counters, which also sets the memory requirements. The graph in Fig.~\ref{fig:relErrorVsCSparams} shows the approximation error for different combinations of $b$ and $t$. 

The CS algorithm provides a tolerable error rate as long as $t\times b \ge 64\cdot 10^6$, except for the case $(t,b) = (4, 16*10^6)$ which has too small of a $t$, causing a high rate of false positives; we will provide more details in the next paragraph. To better understand the spectrum of possible error rates, we consider the rates at rank $400000$, where the frequencies of the cells are already quite low. For all combinations of the parameters, the algorithms start losing a significant fraction of correct heavy hitters near that value. The following table shows these error rates as a function of $b$ and $t$. 

\begin{center}
 \small
 \begin{tabular}{||c c c c c c c c c||} 
 \hline
 t              & $16$ & $12$ & $8$ & $16$ & $12$ & $16$ & $8$ & $12$ \\ 
 \hline
 b $/10^6$ & $16 $ & $16$ & $16$ & $8$ & $8$ & $4$ & $8$ & $4$ \\
 \hline
 Error          & $0.27$ & $0.34$ & $0.42$ & $0.47$ & $0.51$ & $0.64$ & $0.67$ & $0.72$ \\
  \hline
\end{tabular}
\end{center}

Algorithms with a similar space usage ($\propto t\times b$) have a similar error rate, but the solution with the larger number of rows is generally somewhat better. This can be easily explained using a theoretical argument: the largest portion of the relative error depicted is due to losing true heavy hitters; this happens due to the fact that the algorithm finds ''fake'' heavy hitters, and those push the true heavy hitters with a smaller frequency out of the top group. A fake heavy hitter can appear only if it collides with some other true heavy hitter in at least half of the rows. 
Thus, the expected number of collisions can be computed as a total number of different items $n$ times the probability to have collided with at least one heavy hitter, which is $k/b$, and then this should happen in $t/2$ independent experiments. Therefore expected number of collisions is $n(k/b)^{t/2}$. 
From that dependency, we can see that under fixed $t \times b$, the larger value of $t$ is always better.  
For example, if we want to find only one heavy hitter and minimize the space usage which is proportional to $t\times b$, then the most efficient way is to take $b = 2$ and $t = c\log n$. 
However, this would force us to increment or decrement $O(\log n)$ counters for each update, which is much slower, $O(\log_b n)$ with $b>>2$. Theoretically running Count Sketch with $t = 8$ would be twice slower than with $t'=4$, due to the need to compute twice more hashes and increment twice more counters. In practice, we saw almost the same, mostly due to the fact that computing hashes and updating counters takes around $75 \%$ of the total running time in the ''GPU hierarchy'' implementation.   
From these examples, we can understand the nature of the space versus time trade-off, and we can see this behavior in the graph and in the table for the pairs $8 \times 16\cdot 10^5$ and $16 \times 8\cdot 10^5$, $16 \times 4\cdot 10^5$ and $8 \times 8\cdot 10^5$ and others. Note that increasing both $b$ and $t$ will provide better approximation and lower false positive rate, however increasing $t$ would significantly push time performance and space (if we will keep $b$ fixed) up, while increasing $b$ will not influence the time performance but still push the memory usage. In all our experiments we were limited by the memory of the GPU, which for both devices was only 8GB. 

\subsection{Evaluation of the model:}

Here we will evaluate the quality of the model for two specific problems: finding halos and the analysis of excursion sets. To do so, we will try to solve the problem using Count Sketch and its ability to find top $k$ densest cells in the simulation box. 
\begin{figure}[h]
            \includegraphics[width=8cm]{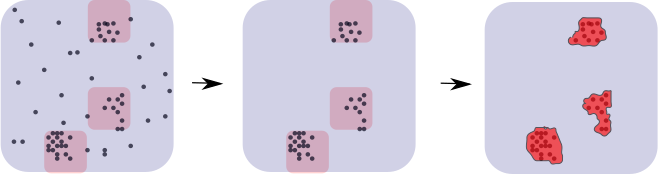}
            \centering
            \caption{Finding halos from heavy cells exactly by running any offline in-memory algorithm on the subset of particles belonging to the top heaviest cells.}
            \label{fig:cells2halo}
\end{figure}

In \cite{liu2015streaming} authors showed a simple solution for using the heavy cells to find heavy clusters by making a second pass through the data set and storing locally the particles which belong to one of the heavy cells. This is possible because the number of particles in those heavy cells is much smaller than that of the entire data set, and we can even store them in main memory. This implies that any traditional in-memory algorithm can be applied offline. This scheme is illustrated in Fig.~\ref{fig:cells2halo}. In this paper, we will not repeat the entire chain of this computation, but will simply check the number of halos contained in the top $k$ cells. 

It turns out  that to find the centers of the $10^5$ most massive halos we need to find the 
$\sim 1.8 \cdot 10^5$ heavy cells, i.e. the centers of the top $10^5$ most 
massive halos are contained in the top $\sim 1.8 \cdot 10^5$ heavy cells. 
Then running an offline in-memory halo finder, such as Friends of Friends~\cite{davis1985evolution} or any other halo finder of 
choice~\cite{knebe2011haloes}, on the particles located only 
inside the top $1.8 \cdot 10^5$ heavy cells (and it's immediate neighbours) 
will provide us with more precise halo centers and mass distribution for each halo. 
We emphasize on the fact that in current manuscript we find only the centers of the haloes 
and leave finding actual borders and mass distribution for future research. 
Hence the streaming approach can be considered as a sieve step, 
allowing us to efficiently remove the particles which are not in the largest halos 
from further consideration. The resulting filtered data set is significantly smaller 
in size, thus one can apply offline algorithms. In \cite{liu2015streaming} we showed 
how to find $10^3$ largest halos while working with a data set of size $10^{9}$: 
find the top $2\cdot 10^3$ heavy cells and run an offline algorithm on the particles 
that are located only inside the heavy cells.

Applying the same approach to find the top $10^6$ heaviest cells in the Millennium
data set containing $10^{10}$ particles would be challenging, but still manageable: 
\begin{enumerate}
\item top $10^6$ haloes contain $\sim 3.8 \cdot 10^9$ particles $= 45$ GB;
\item top $10^5$ haloes contain $\sim 2.5 \cdot 10^9$ particles $= 31$ GB;
\item top $10^4$ haloes contain $\sim 1.4 \cdot 10^9$ particles $= 16$ GB;
\item top $10^3$ haloes contain $\sim 0.8 \cdot 10^9$ particles $= 9$ GB;

\end{enumerate}
Thus we indeed can afford to run offline in-memory halo finder and locate $\sim 10^5$ 
haloes on a desktop or a small size server. At the first glance, it seems that the memory 
gain is not significant: initial dataset weights $\sim90GB$, i.e. for the top $\sim 10^5$ 
haloes the gain is at most factor of $3$ (factor of $5$ for the top $10^4$), 
at the cost of  introduced approximation and non-zero probability of failure. 
However, initial dataset does not provide an option of running offline halo finder 
sequentially in several passes on the machine with very low memory usage. 
Our filtering step provide an opportunity to run it on the machine with just $2$GB of memory,
the algorithm will require make more passes over the data, i.e. to find the top $10^4$ 
haloes one will need to make $\sim 10$ passes while working under $2$GB memory restriction.

We should take into account that number of particles in each halo is growing with 
the size of the data set. Additionally, using  $10$ times larger top will 
significantly increase the total number of particles one need to store in 
the memory, while applying offline algorithm. 
Hence, finding the top $10^5$ haloes on the Millenium XXL dataset with 
$3 \cdot 10^{11}$ particles is less feasible as a low-memory solution: 
\begin{enumerate}
\item top $10^6$ haloes contain $\sim 31 \cdot 10^9$ particles $= 372$ GB;
\item top $10^5$ haloes contain $\sim 9 \cdot 10^9$ particles $=108$ GB;
\item top $10^4$ haloes contain $\sim 2 \cdot 10^9$ particles $=24$ GB;
\item top $10^3$ haloes contain $\sim 0.4 \cdot 10^9$ particles $=4.8$ GB.
\end{enumerate}
From the list above we can see that to keep everything on the small server, proposed 
approach can help to find at most top $10^4$ haloes in one extra pass or $10^5$ haloes in 
$\sim 8-10$ passes, which we state as a result in the current paper and keep the 
further improvement as a subject for future investigation. Note that compression level for 
the top $10^5$ particles is $33$ times ($148$ times for the top $10^4$ haloes). However
requirement to make more than $2$ passes and utilize $\sim24$ GB on the second pass is 
very restrictive and better techniques should be proposed for after-processing. 
Among the most straightforward solutions are sampling and applying streaming 
approach hierarchically for different cell sizes.

\begin{figure}[h]
\includegraphics[width=9cm]{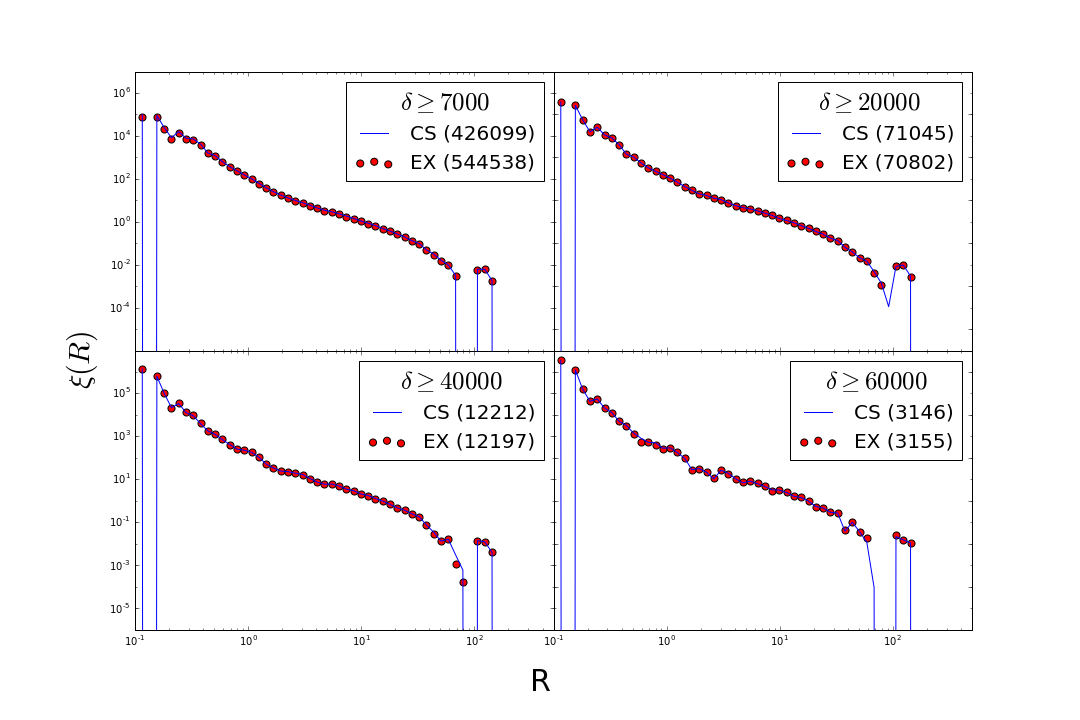}
\centering
\caption{Comparison of the 2-point correlation functions of excursion sets determined using the exact counts and the Count Sketch results for 4 over-density levels. The numbers in parentheses indicate the number of cells that was found and used i the calculation of $\xi$. Clearly the results of applying the spatial statistic to the Count Sketch result is equivalent to that of the exact counts. 
The radius $R$ is in the natural, co-moving units of the simulations, $Mpc/h$.}
\label{fig:2ptfunctions_cs_vs_ex}
\end{figure}

\begin{figure}[h]
\includegraphics[width=8cm]{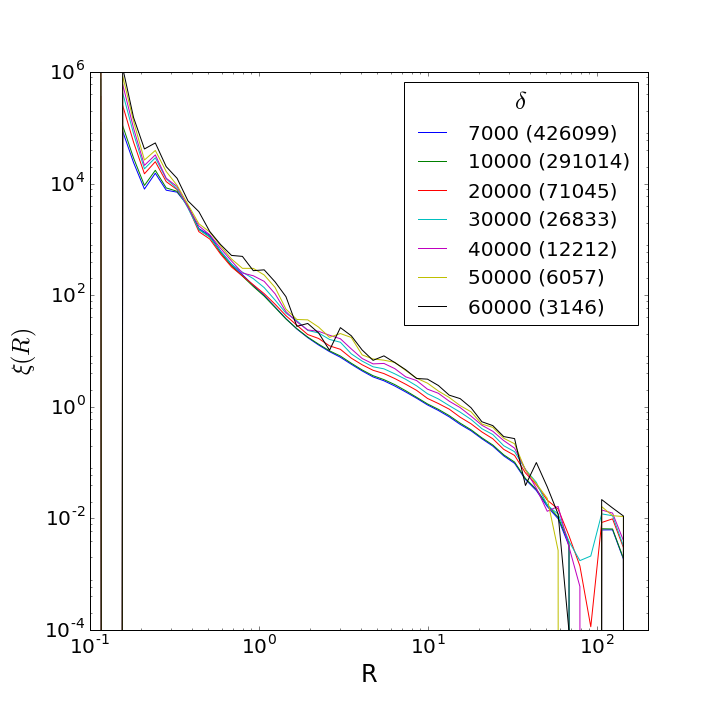}
\centering
\caption{Two-point correlation functions of excursion sets, defined as sets of cells with a certain lower limit on the over-density. In this plot the results of the count-sketch algorithm for detecting heavy-hitters is used to determine the excursion sets. The number next to the line segments in the legend gives the over-density,  the numbers in parentheses indicate the number of cells at that over-density. The radius $R$ is in the natural, co-moving units of the simulations, $Mpc/h$.}
\label{fig:2ptfunctions_cs}
\end{figure}

As described in the introduction, a connection can be made between the heavy hitters in the collection of grid cells and excursion sets of the density field. We want to determine spatial clustering properties of these over-dense cells and determine if the algorithm by which this set is determined has an influence on the spatial statistics. To do this we have extracted the locations of heavy hitter grid cells in the Count Sketch result and determined their clustering properties using the two-point correlation function $\xi(R)$ \cite[e.g.][]{peebles1980large}. We compare this to the 2-pt function calculated on the cells in the exact excursion set. Adopted cell size is $0.1$ \mpch. As the results in Fig.~\ref{fig:2ptfunctions_cs_vs_ex} show, for the over-densities that can be reliably probed with the streaming algorithm the exact and Count Sketch results are indistinguishable. The main deviations are due to discreteness effects for the smaller high-density samples.
As an aside, we note that in Fig.~\ref{fig:2ptfunctions_cs}, the higher-density cells cluster more strongly than the lower-density cells, as expected \cite{kaiser1984spatial}.

\subsection{Millennium XXL}
Running on the Millennium dataset, we could still find the ``top-$k$'' densest cells exactly using quite moderate time and memory. Even when the full density grid was too large to fit in memory, we could make multiple passes over the data and determine parts of the grid. Those experiments are necessary for evaluating the precision of the output from the randomized algorithm, but on our medium sized server they take about a day to complete. 

In this section, we describe an experiment on the results of the Millennium XXL simulation \cite{angulo2012scaling}, which contains around $300$ billion particles, and hence is $30$ times larger than the Millennium dataset. Its box size is $3$ Gpc and we will use a cell size of $0.2$ Mpc. Thus our regular mesh would contain  $\sim 3.4\cdot10^{12}$ cells and need $13.5TB$ of RAM to be kept in memory completely (using 4-byte integer counters). While this is beyond the means of most clusters, our algorithm will be able to solve the problem with a memory footprint under $4$GB while keeping the lapse time under an hour. 

Before we describe some technical details of the experiment, we need to clarify the process of evaluation, as we are now not able to produce exact counts in a reasonable amount of time. Hence, we consider only the following two ways for evaluating the accuracy of our results:
\begin{enumerate}
    \item \textbf{From the size of the exact counts}\\
        While we can not determine the top-$k$ most dense cells precisely, we can still make a second pass over the data and maintain counters for some subset of cells. We will use this opportunity to evaluate the approximation error from Count Sketch, but only for those particles which were output by the algorithm. Note that this verification is not as reliable as the one used earlier in this paper because it does not catch any false negative items, i.e. the items which are supposed to be in the top-$k$, but were lost by the algorithm. But this way we can evaluate the approximation error, and get a preliminary estimation of the false positive rate.  
        
    \item \textbf{From astrophysics} \\
        Running a simulation with a larger number of particles provides us with more stable quantities. While we do not have any way to verify them precisely, we know that the spatial statistics should be more or less close to those from smaller size simulations. This evaluation is more qualitative than quantitative, but it will definitely be alarming if serious deviations are present.
        
\end{enumerate}

\begin{figure}[h]
\includegraphics[width=9cm]{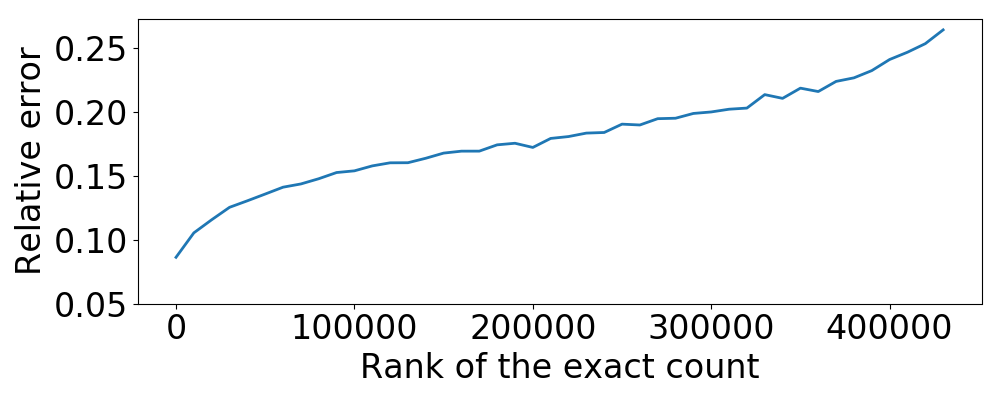}
\centering
\caption{Relative error for the counts in output of Count Sketch algorithm for Millennium XXL dataset.}
\label{fig:XXLrelError}
\end{figure}

We ran the "GPU hierarchical" version of the Count Sketch. We then made a second pass over the dataset where we determined the exact counts, restricted to the cells found in the first pass. While we do not know the cutoff frequency for the top-$k$, we can still approximately estimate the false positive rate: if all cells in the top-$k$ output by Count Sketch have frequencies larger than $1700$, then every item with a true frequency less than $100$ would be considered as false positive. Initially, we set the same number of counters in the Count Sketch table as before: $16$ rows and $10^7$ columns. However the result was quite noisy and had a very high rate of false positives: around $60000$ had a frequency lower than $100$, while the top-$k$ frequency cutoff is around $1700$. Then we ran Count Sketch with $24$ rows, and the number of collisions dropped accordingly to approximately $800$. The graph depicting relative error of the Count Sketch can be observed in Figure \ref{fig:XXLrelError}. It is evident that approximation error is more than twice that of the experiment with the Millennium dataset (refer to the Figure \ref{fig:relError}). This can be explained by the size of the dataset, as long as algorithm's guaranteed approximation error is $\varepsilon \ell_2$, then with  $\ell_2$ for the XXL dataset the error is significantly larger. If the further astrophysics analysis will require better approximation error we can increase the width of the Count Sketch table. 

\begin{figure}[h]
\includegraphics[width=9cm]{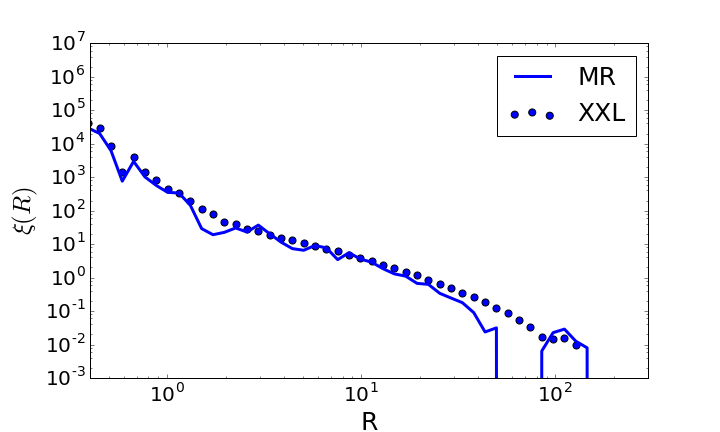}
\centering
\caption{Comparison of CS 2-pt correlation function for excursion sets in 0.2~Mpc cells with $\delta\geq 20000$ for the XXL (dots) compared to the exact result for the Millennium run. The two results are compatible with each other, with deviations explained by discreteness effects in the much sparser Millennium result. The radius $R$ is in the natural, co-moving units of the simulations, $Mpc/h$.}
\label{fig:2pt_MRvsXXL}
\end{figure}

In Fig.~\ref{fig:2pt_MRvsXXL} we compare the two-point correlation function for all cells with $\delta\geq 20000$ for both Millennium and Millennium XXL. For the Millennium XXL result we use the Count Sketch, for the Millennium we use the exact over-densities, both in cells of size 0.2~Mpc. The XXL has a volume that is $216\times$ the volume of the Millennium run and hence much better statistics. Nevertheless, the results are compatible with each other.

We ran the experiment on the small server with the following characteristics: Intel Xeon X5650  @ 2.67GHz, 48 GB RAM, GPU Tesla C2050/C2070. Our I/O was based on 4 Raid-0 volumes of 6 hard drives each. The total time for the I/O is 30 minutes. Due to the fact that I/O is implemented in parallel, if the time of the algorithm is higher than I/O, then I/O is "for free", this happens due to the fact that we can read a new portion of the data from the disk, while the GPU is still processing the previous portion. Our algorithm time on the Tesla card is 8 hours. In contrast, on the GTX1080, the estimated running time is expected to be less than an hour, which we calculated by running a small portion of the data. However, we were not able to carry out the entire experiment on the GTX1080, lacking a server with this card, and parallel I/O with a significant amount of storage. 

For comparison, had we calculated the exact density field on a grid we would have required about 13.5TB of memory. Alternatively, on the machine with 48 GB RAM, we'd need about 280 passes over the data to calculate the exact field in chunks small enough to fit on the machine. 

\section{Conclusion}
\label{sec:conclusion}
In this paper, we have carried out a detailed investigation of applying streaming algorithms to cosmological simulations. Our first proof-of-concept results, introduced in \cite{liu2015streaming}, uncovered the ability of so-called ''heavy hitter'' streaming algorithms to determine density statistics, making only one pass over the data. In the current paper, we pushed the limits of these algorithms toward datasets with sizes up to $\sim 10^{12}$ particles, while still keeping all computations on a single server, or in some cases, even a desktop. To make this possible, we implemented the Count Sketch algorithm in a batch streaming setting and ported it to a graphics processor (GPU) using the CUDA environment. This approach significantly improves the time performance while using much less memory, enabling the possibility of processing very large datasets. 

We have benchmarked several implementations, varying time, precision, and memory usage. We conclude that GPUs offer a perfect infrastructure for supporting the batch streaming model. Note that in the current project, while all experiments were carried out on a single GPU, we did not change the Count Sketch data structure. Thus, two or more sketches computed on different nodes, if merged, will approximate the cell counts for the combined stream of updates. Therefore, this approach can be used in distributed settings, where each node will have its own stream of updates and its own data sketch, and at the end all the sketches can be summed to find the heaviest cells. An implementation of this algorithm on distributed storage, using several GPUs, is crucial due to IO being the main bottleneck and will be considered in future work. Additionally, we will investigate the application of other classic streaming algorithms in a batch streaming model, on the GPU. Among other future directions, we are considering structure finding in 6D space, where each particle is described by its velocity and location; we are also considering hierarchical sketch-based clustering, to find the top-$k$ heaviest cells in meshes of different sizes in parallel.

Though the emphasis in this paper is on the technical application of these streaming algorithms in a new context, we showed, where possible, that these randomized algorithms provide results consistent with their exact counterparts. In particular, we can reproduce the positions of the most massive clusters and the two-point correlation function of highly non-linear excursion sets. The nature of these algorithms currently precludes the possibility of sampling the full density field or the full halo multiplicity function, though we are working on algorithms to at least approximate those statistics. 

\section{References:}
\bibliographystyle{plain}
\bibliography{references}
\end{document}